**Cornelia Victoria Anghel-Drugărin***

# PROIECTAREA PRIN INTERMEDIUL UNUI SOFTWARE DEDICAT A INSTALAȚIEI ELECTRICE PENTRU UN IMOBIL

# A SOFTWARE DESIGN THROUGH ELECTRICAL SYSTEM FOR A BUILDING

**Rezumat**

Proiectarea asistată de calculator a sistemelor de iluminat realizează dimensionarea unor instalații noi de iluminat cât și verificarea celor existente, atât pentru sistemele de iluminat interior cât și pentru sistemele de iluminat exterior.

În proiectarea realizată s-a utilizat programul dedicat DiaLux versiunea 4.11.

**Abstract**

Computer aided design of lighting systems made new installations of lighting dimensioning and verification of existing lighting systems for both indoor and outdoor lighting systems.

The design made to use program dedicated to DiaLux version 4.11.

## 1. INTRODUCTION

Computer aided design of lighting systems involving specific knowledge and expertise in building design technology facilities (for power and protection of technological devices).

Commencement of the design must be based on the design theme developed based on a study of necessity and opportunity of investment.

Achieving efficiency requires completion of at least the following design criteria:
- introducing the most modern projects and solutions-technological processes in the world of science and technology;
- optimal sizing systems and equipment;
- choice of solutions involving reduced consumption of materials;
- inclusion in the draft assemblies standardized and standardized;
- adopting those technological solutions that provide rational use of energy resources;
- choosing green solutions, aesthetics, maintenance, ergonomics and high operational safety;
- adopting those solutions that increase the productivity and to improve processes.

The main task of the designer is to determine how best to achieve low-voltage electrical installations.

In the area of houses and office jobs, one of the most important conditions is related to providing comfort and ergonomics that is achieved through the provision of adequate facilities and aesthetic situations which contribute to peace and rest and relaxation on the one hand and to focus on the other. Also, the current information transmission that almost permanently to be put into doeth went to work in a dose that work-related effort so that using computers and telephony to expand close communication throughout the day. Thus we need almost anywhere, so by default for administrative units have access to the technical means necessary for this purpose: telephone, television, Internet, etc.

An important aspect to be considered in residential complexes is based on the costs and depending on equipment or payback time.

2. APLICATIONS

2.1. Initial data about the project

We considered a duplex house, consisting of living room, large type 1 room, type 2 large room, small room type 1 and small room type 2, bathroom, lounge, kitchen.

For the considered property type and number of units for apartments rooms and hallways analyzed are shown in the table below:

*Tabel1.* Type and number of units for apartments rooms and hallways.

| No | Area | Device/ Light source [buc] | | | | |
|---|---|---|---|---|---|---|
|  |  | Lamp | Monopolar intreruptor | Bipolar Swicher | Head Scale Swicher | Monophasic sochets |
| 1 | Hallway access | 1 | 1 | - | 2 | 1 |
| 2 | Hallway1 ground floor | 2 | - | 1 | 2 | 1 |
| 3 | Hallway2 ground floor | 2 | - | 1 | 2 | 1 |
| 4 | Hallway1 level1 | 2 | - | 1 | 2 | 1 |
| 5 | Hallway2 level1 | 2 | - | 1 | 2 | 1 |
| 6 | Porch | 1 | 1 | - | 1 | 1 |
| 7 | Stairs ground level to level1 | 2 | - | - | 2 | 1 |
| 8 | Stairs | 2 | - | - | 2 | 1 |
| 9 | Central corridor floor | 2 | - | 1 | - | 1 |
| 10 | Garage | 2 | - | 1 | - | 1 |
| 11 | Storage | 2 | - | 1 | - | 1 |
| 12 | Bedroom 1 | 1 | 1 | 1 | - | 3 |
|  | Small room t 1 | 2 | - | 1 | - | 2 |
|  | Living type 1 | 3 | - | 1 | - | 3 |
|  | Bathroom 1 | 1 | 1 | - | - | - |
|  | Kitchen type 1 | 2 | - | 1 | - | 2 |
| 13 | Bedroom type 2 | 2 | - | 1 | - | 3 |

|    | Small room t 2   | 2 | - | 1 | - | 2 |
|----|------------------|---|---|---|---|---|
|    | Big room tip 2   | 3 | - | 1 | - | 3 |
|    | Bathroom tip 2   | 1 | 1 | - | - | - |
|    | Kitchen tip 2    | 2 | - | 1 | - | 3 |
| 21 | Bedroom 3 tip 1  | 2 | - | 1 | - | 2 |
|    | Bedroom3 tip 2   | 2 | - | 1 | - | 2 |

The main parameters and coefficients necessary to calculate the lighting system (measured or taken from the rules) stated:
- the reflection coefficient;
- for the ceiling: concrete materials plastering or matte paint (possibly applied to plasterboard) pt = 0.5;
- for the Walls with similar characteristics, possibly coverings: ρp =0.5 or pp = 0.3;
- ambient lighting for living spaces are $E_{med}$ = 50 Lx for rooms, for hallways $E_{med}$ = 75Lx, for main spaces $E_{med}$ = 20 Lx and $E_{med}$ = 100 LX for annexes and offices.

## 2.2. Applying the coefficient of Use method

For dimensioning the lighting of the rooms and spaces with special lighting, strict requirements and utilization coefficient method is applied.
This method is used for symmetric luminaries of the same type placed at a height equal to the plane manner.
Utilities of the rooms and hallways are considered the height $h_u$ = 1,6 m
located approximately at eye level for the displacement situation.
Luminaries with light sources are placed symmetrically keeping the rooms and building elements to furniture, ensuring a certain aesthetic in accordance with the wishes of the beneficiary. Distance suspension and fixation is given room size and length measurements resulting from power conductors.
The necessary data are summarized in tables.
Consider the following hypotheses:
- reflection coefficients of the ceiling and walls are considered constant;
- reflection coefficients of furniture is ignored.
The coefficient of utilization is the ratio of flow u manner that it falls on the surface of $\Phi_u$ and the total flux emitted by the source $\Phi_t$:

$$u = \frac{\Phi_u}{\Phi_t} = \frac{E_{med} \cdot S_u \cdot K}{N_c \cdot n \cdot \Phi_l}$$

where:
- $E_{med}$ is average luminance imposed on useful plan;
- $S_u$= A·B represents horizontal area of the space to the useful plan;
- K is depreciation coefficient that takes into account the decrease over time of lamp lumen maintenance intervals and lighting installation;
- $N_c$ is the number of lighting in a room that will be established according to the preference beneficiary;
- n is the number of lamps is fitted luminaries lamp;
- $\Phi_l$ is the luminous flux of a lamp (500...600 lm for our study case).

Utilization coefficient is an overall yield of actual use of the flow of light sources useful plan.

As the height h over the useful plan is higher, such as the solid angle which includes the surface of the falling flow capacity is lower and the rate of use is lower.

$$h = H - (h_u + h_a)$$

The size of the previous formula have the following meanings:
- H is the total height of the rooms (results of measurements and presented and summarized in the table)
- $h_u$ is the utilities room height, which is noted $h_u = 1,6$ m in our study case
- $h_a$ is the distance from the top of the suspension or fitting room lighting;
$h_a = H - h$, where h is the height of suspension or mounting luminaries lamp established based on user requirements
The surface of the rooms in the useful plane is calculated as follows:
- for the large room type 1:
$S = A \cdot B = 4,8 \cdot 4,2 = 19,2 m^2$
- for hallway landing:
$S = A \cdot B = 5,8 \cdot 1,6 = 22,2 m^2$.

### 2.3. Implementation with CAD for light systems DiaLux, version 4.11

DIALUX 4.11 version, European software program, is dedicated to the design of lighting systems used by major manufacturers in the industry. The software can be free downloaded from SC Elba siteTimişoara.

The software is applicable to: interior lighting system (offices, warehouses, commercial), external light (commercial spaces, monuments, parks, sports fields), or on the road (streets, intersections).

The program is useful for sizing interior lighting and areas having the following functions and structure:
- New lighting project type
- Characteristics of the room or area,
- User choice reflectance in a room
- Selection of light sources and lighting-database;
- Location lighting;
- Recommended lighting levels for various indoor destinations;
- Surface and calculation points;
- Calculations, average and spot lighting;
- Save Data;
- Print results.

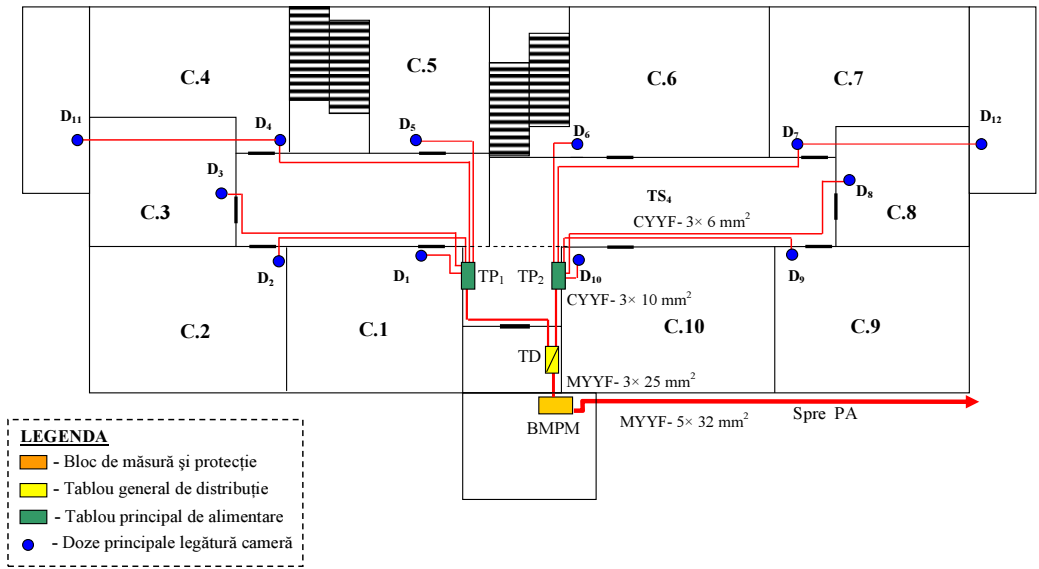

**Figure 1.** Sizing wiring diagram in the building.

The scheme of our building are 0.4 kV electrical install. Also on the schedule is noted and sectioning system. The whole plant is placed under the plaster hoses. The wires used were dimensions 5X32 mm$^2$ 3x25 mm$^2$ 3x10 mm$^2$ and 3x6 mm$^2$ and power plants belong.

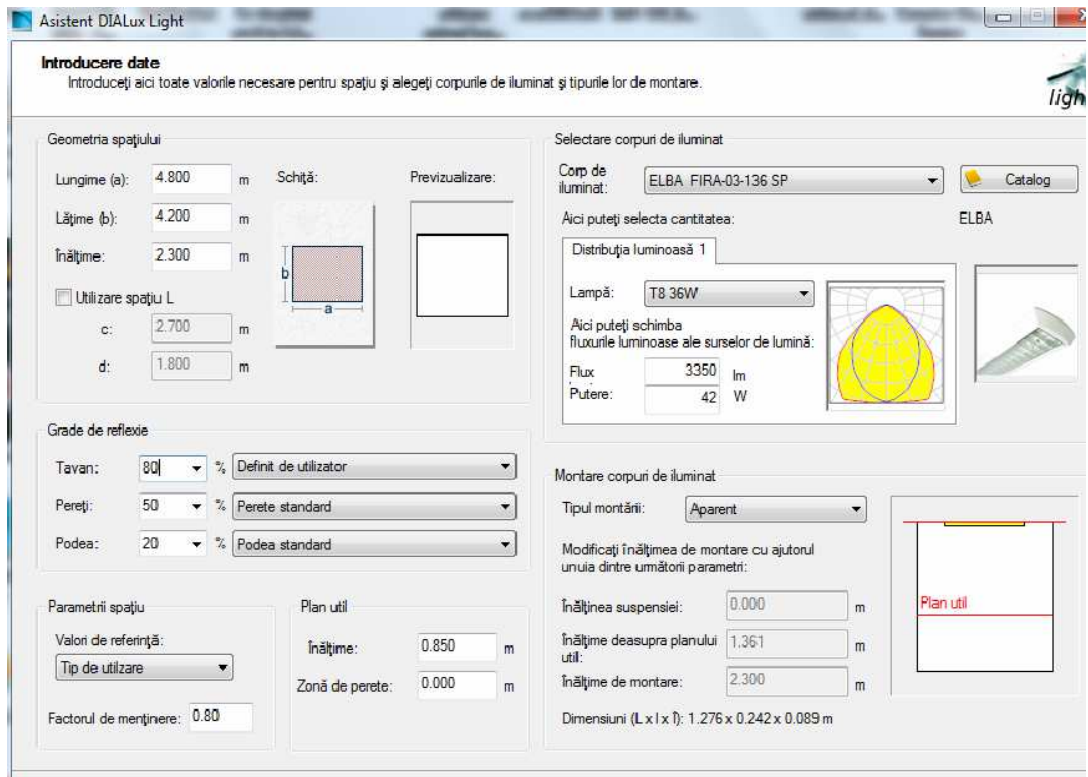

**Figure 2.** Inserting values necessary for space and choice of mounting bodies and their types.

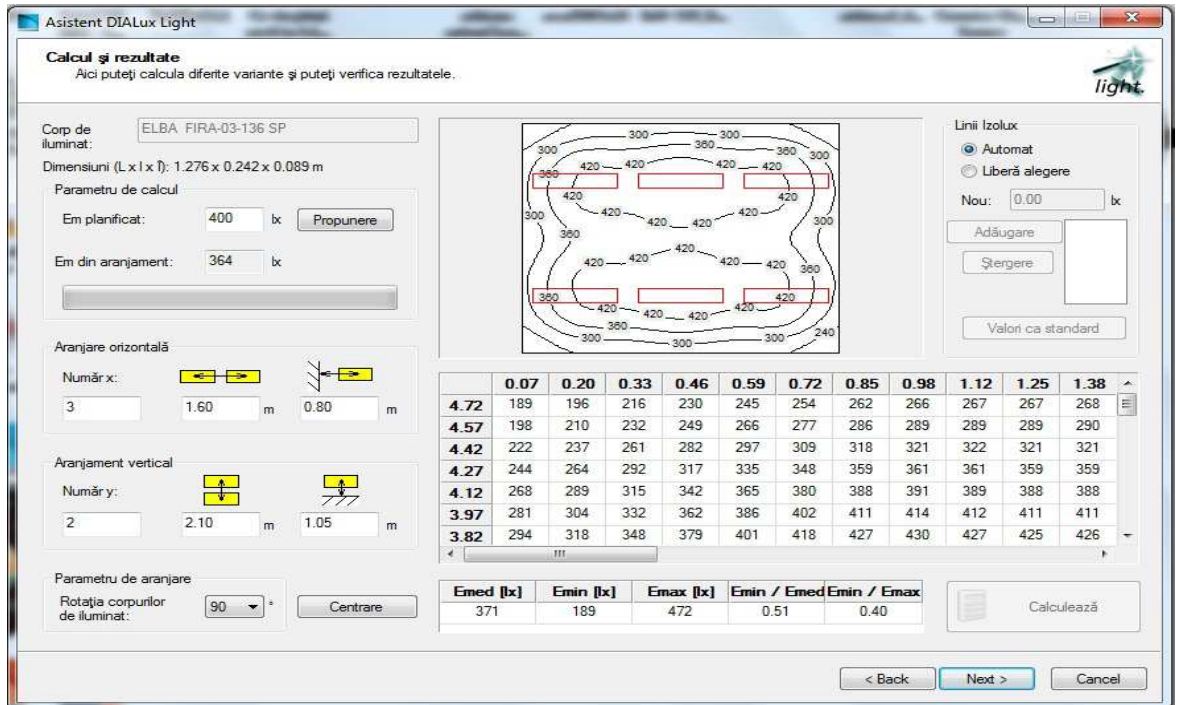

**Figure 3.** Calculating different settings of lighting and checking results.

Finally mutually agreed with the customer, location and placement of lighting outlets are established.

## CONCLUSIONS

Building energy management system uses tracking interconnection means contributing to the development and implementation of intelligent building concept ensuring an optimal energy management thereof.

It was considered a duplex house consisting of living room, large room type 1, large room type 2, small room type 1 and small room type 2, bathroom, lounge, kitchen.

The design of the electrical installation was performed with dedicated software Dialux version 4.11.

**\* PhD Eng**. **Cornelia Victoria Anghel-Drugărin** – "Eftimie Murgu" University of Reşița, Electrical Engineering and IT Department, Computer Science Technology, c.anghel@uem.ro, vicepresident ACM-V Caras-Severin Region, Romania.